\documentclass[a4paper, 12pt, journal, doublecolumn]{IEEEtran}  
\usepackage[colorinlistoftodos,prependcaption,textsize=tiny,color=red!15]{todonotes}
\usepackage{flushend}
\usepackage{multirow}
\usepackage{nameref}

\begin{document}

\title{Mechanical problem solving in Goffin's cockatoos\\-- Towards modeling complex behavior}

\author{Manuel Baum*$^{1,2}$ \qquad Theresa Roessler*$^{3,4}$ \qquad Antonio J. Osuna-Mascaró$^{3}$ \\ Alice Auersperg$^{2,3}$ \qquad Oliver Brock$^{1,2}$ 
\thanks{* These authors contributed equally.}  
  \thanks{$^1$ Robotics and Biology Laboratory, Technische Universit\"at Berlin, Germany}
  \thanks{$^2$ Science of Intelligence (SCIoI), Cluster of Excellence, Berlin, Germany}
\thanks{$^3$ Comparative Cognition, Messerli Research Institute, University of Veterinary Medicine Vienna, University of Vienna, Medical University of Vienna, Austria}
\thanks{$^4$ Department of Cognitive Biology, University of Vienna, Austria}
  \thanks{We gratefully acknowledge funding by the Deutsche Forschungsgemeinschaft (DFG, German Research Foundation) under Germany's Excellence Strategy -- EXC 2002/1 ``Science of Intelligence'' -- project number 390523135. Theresa Roessler was supported by FWF START Project Y1309 to Alice Auersperg.}

}

\maketitle


\begin{abstract}
Research continues to accumulate evidence that Goffin's cockatoos (\textit{Cacatua goffiniana}) can solve wide sets of mechanical problems, such as tool use, tool manufacture, and solving mechanical puzzles. However, the proximate mechanisms underlying this adaptive behavior are largely unknown. In this study, we analyze how three Goffin's cockatoos learn to solve a specific mechanical puzzle, a lockbox. The observed behavior  results from the interaction between a complex environment (the lockbox) and different processes that jointly govern the animals' behavior. We thus jointly analyze the parrots' (1) engagement, (2) sensorimotor skill learning, and (3) action selection. We find that neither of these aspects could solely explain the animals' behavioral adaptation and that a plausible model of proximate mechanisms (including adaptation) should thus also jointly address these aspects. We accompany this analysis with a discussion of methods that may be used to identify such mechanisms. A major point we want to make is, that it is implausible to reliably identify a detailed model from the limited data of one or a few studies. Instead, we advocate for a more coarse approach that first establishes constraints on proximate mechanisms before specific, detailed models are formulated. We exercise this idea on the data we present in this study.
\end{abstract}


\section{Introduction}


One of Tinbergen's famous four questions for the study of animal behavior is about the mechanisms~\cite{tinbergen_aims_1963} underlying behavior. These mechanisms are cognitive processes, reactions to stimuli, or physiological phenomena. It is crucial to understand them in order to understand animal behavior and especially its generality. Generalization capabilities allow animals to solve novel tasks or new variants of previously encountered tasks. It will not only benefit behavioral biology to comprehend the mechanisms underlying generality, it will also benefit approaches to engineering machine behavior, such as in robotics. However, to research and elicit generalization behavior in animals we need to test them in novel and possibly challenging tasks. Such tasks jointly recruit different, heterogeneous processes when an animal behaves and adapts to solve them. As task-directed adaptation can happen individually or jointly in different processes, it is crucial that behavioral analysis also aims to jointly analyze the different processes that may be the basis for adaptation. In this paper we perform such a joint analysis of different adaptive factors in a challenging task. We analyze how engagement, sensorimotor skills, and task-solving strategy change over time as Goffin's cockatoos learn to solve a mechanical puzzle task.      

\begin{figure}\center
\includegraphics[width=\linewidth]{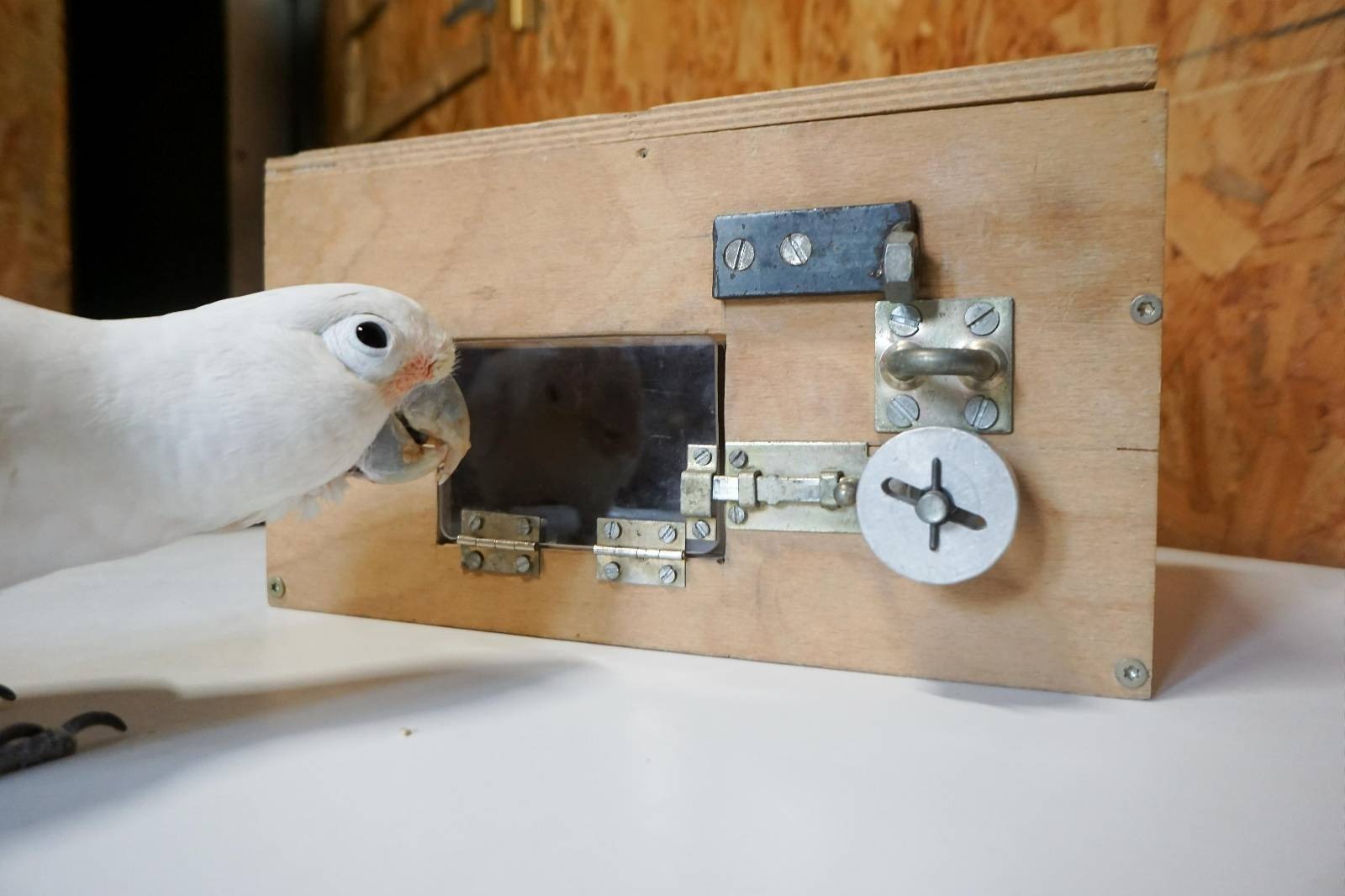}
\caption{Zozo in front of the lockbox -- a baited mechanical puzzle where the reward can only be obtained after a sequence of mechanisms is unlocked. The configuration is as presented to the subjects; window closed, bar in locked position and wheel attached with the opening horizontal so as to be perpendicular to the pin (already slightly rotated by Zozo in this picture).}
\label{fig:lockbox}
\end{figure}

Goffin's cockatoos (\textit{Cacatua goffiniana}) are a species that shows impressively general manipulation and problem solving abilities. They excel at manipulating and combining objects, as well as at haptic exploration~\cite{auersperg_unrewarded_2014,auersperg_combinatory_2015}. Furthermore, they flexibly use and manufacture tools~\cite{auersperg_spontaneous_2012,auersperg_tool_2018,ohara_wild_2021,osuna-mascaro_innovative_2022,osuna-mascaro_flexible_2023} and skillfully solve a wide range of mechanical problems~\cite{habl_keybox_2017,rossler_using_2020}, such as the lockbox~\cite{auersperg_explorative_2013} in Fig.~\ref{fig:lockbox}. Lockboxes are mechanical puzzles where a reward can only be obtained if the subject opens a sequence of mechanical locks. There were no puzzle boxes in the evolutionary past of Goffin's cockatoos, yet they can learn to solve them skillfully. This behavior builds on abilites and species-specific predispositions that have evolved in their natural habitat. Goffin's cockatoos opportunistically forage on a variety of food sources, including ones that are embedded~\cite{ohara_extraction_2019}. The mechanisms enabling such behavioral generality are unknown until now. It is a major motivation for our research to reveal these mechanisms, and to understand what methodologies are required to achieve this research goal.

It poses a significant challenge to uncover the mechanisms that underlie animal behavior and generality, due to at least two major reasons. Firstly, to observe behavioral generality we need to confront animals with challenging, novel tasks. When exploring and learning to solve such tasks, animal behavior can become highly complex. Thus, a model that can replicate all behavioral facets is unattainable -- as such a model would essentially replicate the animal itself. This can be addressed by developing models with a well-defined focus, e.g. on a certain level of abstraction (e.g. Marr's levels of computation, algorithm and implementation~\cite{marr_vision_1982}), or to specific behavioral phenomena. Secondly, behavior results from a multitude of heterogeneous processes that substantially interact~\cite{acerbi_method_2022,bandini_examining_2020,griffin_innovativeness_2016,schierwagen_reverse_2012,biondi_behavioural_2022} and are context dependent (as described for example in \cite{biondi_behavioural_2022,griffin_innovation_2014,griffin_innovative_2016,horn_beyond_2022}). 

These two issues imply conflicting requirements in modelling. On one hand, a model should have limited focus, but ideally it should also comprehensively encompass behavior to account for the interaction between involved processes. Concretely, when Goffin's cockatoos tackle lockboxes, their behavior results from various mechanisms, including those governing their engagement, mechanical skills and their strategy to choose interaction points on the lockbox. As these apsects are not independent, an ideal model should encompass each of these factors and their interaction. However, each of these factors in isolation is already exceptionally complex, such that a joint model with high resolution in each component seems unattainable. 

\begin{figure}\center
\includegraphics[width=0.5\textwidth]{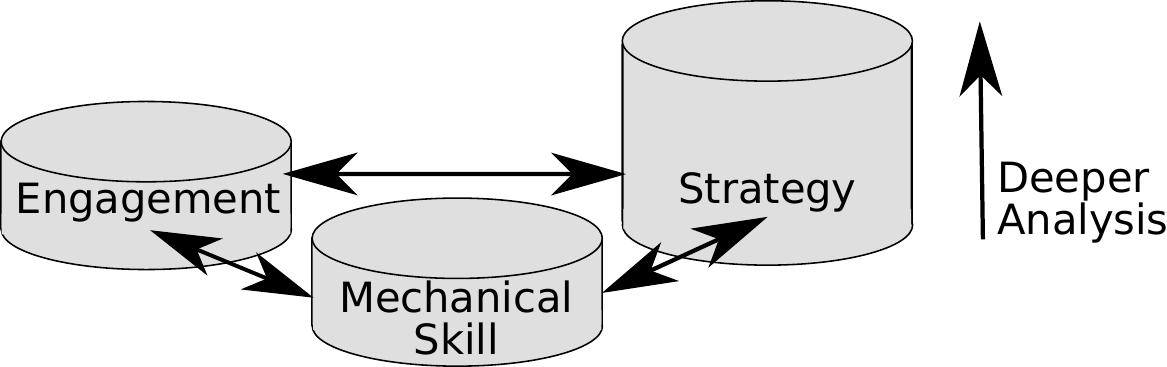}
\caption{Goffin's cockatoos mechanical problem solving behavior depends on their engagement, sensorimotor skills, strategy, and the interaction between these factors. We advocate to jointly analyze these factors, and in this paper provide a more in-depth analysis of their problem solving strategy.}
\label{fig:factored_analysis}
\end{figure}

To address this challenge and to strike a balance, we advocate that, akin to modern statistical approaches, models should simultaneously incorporate these aspects, while placing a more focused emphasis on one of them. As depicted in Fig.~\ref{fig:factored_analysis}, we present a study on lockbox solving in Goffin's cockatoos, where we conduct a comprehensive analysis encompassing engagement, sensorimotor skills, and strategy. However, we set the main focus on the birds' strategy to choose interaction points on the lockbox.  

Through the data presented in this study, our aim is to contribute to efforts aimed at unveilling the mechanisms that underlie lockbox-solving abilities of Goffin's cockatoos. Nevertheless, the profound complexity of situated animal behavior raises the question: to what extent can a single study aspire to reveal proximate mechanisms?

The mechanisms governing situated animal behavior are so complex that typically even multiple experiments will not yield enough data to reliably identify a concrete mechanism. Traditionally, the field of behavioral biology employs experimental designs such as transfer tasks~\cite{shettleworth_animal_2001,auersperg_physical_2017,castro_relational_2017} and control groups ~\cite{ruxton_experimental_2011,quinn_experimental_2002} to glean insights into proximate mechanisms. In both cases, animals are confronted with carefully selected task variations, such that behavioral variations are informative of the behavior's underlying proximate mechanisms.

As behavioral and cognitive biology increasingly conducts such experiments, we gather more information about some mechanistic parameters. Nonetheless, plausible mechanisms capable of explaining lockbox-solving behavior will be complex. Thus, it remains unlikely that we will precisely identify the underlying mechanisms in the near future, even using the aggregated data from multiple studies. It follows that we need to employ an iterative approach to model and identify proximate mechanisms. This approach should not merely accumulate facts, it should gradually converge towards increasingly clear statements about mechanisms. As a step in this direction, we suggest collecting \emph{constraints} on proximate mechanisms. Constraints are statements that narrow down the set of possible mechanisms. While not as sharply defined as concrete model mechanisms, they are more mechanistic than pure observational data. Compared to concrete model mechanisms, they are less likely to be invalidated by future research. As a joint group of researchers from behavioral biology and robotics we realized that such an intermediate level of modeling is essential if we wish to bridge modeling efforts between biological data and implementable algorithmic models that may be executed on robots. After we present experimental data we will list constraints that mechanisms should likely fulfil to explain the lockbox-solving behavior we present in this paper.  


\subsection{Solving a Kinematic Problem From Scratch: The Lockbox}

In this study, we apply the concepts discussed earlier to a practical example. Our study involves three Goffin cockatoos (\textit{Cacatua goffiniana}) facing a unique mechanical challenge known as a "lockbox". This lockbox is a mechanical puzzle that requires a sequence of actions to obtain a specific reward (a palatable nut) at the end. While related work has analyzed similar problem-solving in the same species~\cite{auersperg_explorative_2013}, it focussed on analyzing learned behavior using transfer-tests, and did not postulate constraints on mechanisms. Related robotic work~\cite{baum_opening_2017} presented a mechanistic model that is inspired by biological behavior and solves a robot-sized lockbox, but the utilised mechanisms are not supported by biological data.

In our study we employ a variant of the lockbox in~\cite{auersperg_explorative_2013}, where we investigate how these birds learn to solve the lockbox when a new, previously unencountered lock extends the sequence of mechanical locks. The subjects in our study are captivity-born adults raised and living in a rich social environment, and thus have a complex past history but had no previous experience with the present problem (for a small caveat see below). We recorded and analyzed their behavior from their first encounter with the apparatus in its test configuration to full competence. 

\section{Setting up the Bird Experiments}

\subsection{Subjects \& Housing}

We tested three adult Goffin's cockatoos (two males, Zozo and Muki; one female, Fini. They were 8, 7  and 11 years of age respectively). One male, Zozo, had been initially exposed, but not included, in the previous 5-lock study~\cite{auersperg_explorative_2013} (data collection 2011). At the time he was a 1-year old juvenile and was not able to solve the preliminary steps required to enter the experiment, namely opening the window and sliding the bar (see Pre-Training below). Muki and Fini had not taken part in the 5-lock study. The birds were housed in a group aviary with 13 other Goffins. Their diet consists of a variety of seeds, fruits, vegetables, eggs, and nutritional supplements. The aviary consists of a spacious outdoor compartment ($150 m^2$ ground area, $3-5 m$ height) and an indoor part ($45 m^2$ ground area, $3-6$ m height). The latter is heated to $20^\circ C$ during winter. Testing was conducted individually in an adjacent, visually occluded  test compartment ($9 m^2$, $3m$ height). All birds regularly take part in behavioral studies, in which they have to extract food from different puzzle boxes (see for example: ~\cite{auersperg_goffins_2016,auersperg_tool_2018,beinhauer_prospective_2019,habl_keybox_2017,lambert_goffins_2021,laumer_can_2017,osuna-mascaro_innovative_2022,rossler_using_2020}).

\subsection{Ethics}

The experiment was  not invasive and was therefore not classified as animal experiments in accordance with the Austrian Animal Experiments Act (TVG 2012). The task itself used an appetitive protocol based on the parrots' interest in the rewards and the task. The birds were not food deprived prior to the experiment. All the birds were hand-raised, derived from European breeders, and have full CITES certificates. They are also officially registered according to the Austrian Animal Protection Act (§ 25 - TschG. BGBl. I Nr. 118/2004 Art. 2. 118) at the district's administrative animal welfare bureau (Bezirkshauptmannschaft St. Pölten Schmiedgasse 4-6, 3100; St. Pölten, Austria).
  
\subsection{Apparatus}\label{sec:setup:apparatus}
We used part of the lock apparatus described in~\cite{auersperg_explorative_2013}, termed the “lockbox”. It consisted of a wooden, rectangular box with a transparent acrylic window (see Fig.~\ref{fig:lockbox}). The window could be blocked with a sequence of mechanical devices. In the present experiment, (1)  access to the food reward was impeded by the door that was blocked by a bar that had to be pushed to the right to open the door. Moving the bar could only be done after (2) displacing a wheel forward on its axis, but shifting the wheel was only possible after (3) a partial rotation, so that a slot in the wheel was aligned to a pin in its axis (see Fig.~\ref{fig:lockbox}; Supplementary Video 1). Birds were first pre-trained with the opening mechanisms of the door and the bar. The experimental treatment started when the wheel was first attached. The wheel's slot was perpendicular to the pin at the beginning of each session.

\subsection{Procedure}

\subsubsection{Habituation}

Like most parrot species, Goffin's cockatoos are both highly explorative and neophobic~\cite{greenberg_ecological_2001}\cite{ohara_temporal_2017}. To reduce neophobic reactions towards the lockbox we first habituated the birds to the apparatus: We presented it with sunflower seeds (medium quality rewards) on top and around the lockbox and placed a piece of cashew nut (a high quality reward) inside the box while leaving the window open. This continued until they willingly fed on the cashew without any signs of distress. Additionally, and to reduce neophobic reactions towards the wheel, subjects were fed sunflower seeds from on top of the wheel but when the latter was detached from the apparatus and outside the experimental room, in the group aviary.

\subsubsection{Pre-training}

The lockbox was presented to the subjects with a piece of cashew in the door compartment, which was already opened at this stage. Once they readily took the cashew, the door was gradually closed in repeated presentations, until the birds were skilled in opening it. Subsequently, the door was fully closed and blocked by the bar. From that stage onwards, the subject had to push the bar to the right before being able to open the door. During this training, the experimenter guided the bird's attention to the contact point and presented how the mechanisms worked. These pre-training sessions were conducted until each bird reliably shifted the bar and opened the door within 1 minute and without interference by the experimenter. 
One bird, Fini, showed aversive reactions in the first test session, when the wheel was presented for the first time attached to the lockbox. After 10 minutes, we detached the wheel and placed it  20 cm apart, baited with a seed. Then we reattached it and resumed testing to complete session 1. 

\subsubsection{Test}

In test sessions the configuration was as presented in Fig.~\ref{fig:lockbox} and described in Sec.~ref{sec:setup:apparatus} (see Supplementary video 1). Sessions lasted 15 min or until the reward was reached, whatever was first. We conducted 12 sessions per bird in February - March 2018, during which all three birds became competent in completing the sequence within a session. Two of the birds completed 17 additional sessions in September – November (sessions 13-29; the third subject lost motivation to participate). 
During test sessions the experimenter 
sat next to the birds, wearing mirrored sunglasses and was avoiding lateral head movements. Her only interventions were to occasionally tap on the centre of the table to draw the subject's attention to the task or reposition the bird back onto the centre of the table, if it flew off.
Just before the additional sessions (13 - 29) the birds were fed five small cashews from inside the open box and once from the wheel while detached from the apparatus, to re-acquaint them with the system. 
One subject, Muki, successfully removed the wheel in session 7 but then failed to shift the bar in the remaining 12 min of that session. This subject received two sessions that started without the wheel, namely continuing from the state it had reached in session 7 – until it successfully solved the lockbox in the second of those sessions.

\subsection{Behavioral Coding}

To allow full sight of the behaviors, test sessions were recorded with 2 video cameras and the behavior annotated using the open-access software BORIS (Behavioral Observation Research Interactive Software; version 7). We coded the duration of each session and all defined actions taken towards the lockbox (i.e. physical contacts with each lock, see detailed ethogram in Supplementary).

\subsection{Data Analysis and Implementation}
We performed our data processing and visualisation using custom python scripts. The specific steps of analysis are explained in the next section. Implementations as well as data have been uploaded\footnote{https://osf.io/jqutx/?view\_only=66f513d6748546a195d9ed9f978f4178}. As a first step these scripts convert the data in our BORIS annotation project files to pandas (v 1.1.2) data-frames.  We then use numpy (v 1.22.3) for numerical operations on the data and finally plot the data using matplotlib (v 3.1.2) and seaborn (v 0.11.0). 

\section{Goffin's Cockatoos Adaptation in the Lockbox Task}\label{sec:analysis}


When animals learns to solve a challenging task, like the lockbox, behavioral adaptation may happen in different regards. Lockbox solving performance hinges on three key factors: \emph{a)} the ability to choose the right action based on the current task state, \emph{b)} proficiency in executing that chosen action on the lockbox, and \emph{c)} the animal's level of engagement or involvement with the task. In the following discussion, we break down and define these three aspects, highlighting how they evolve individually during adaptation. However, since these aspects are likely interconnected, it's crucial to consider them collectively. We demonstrate that, based on our definitions, the combined effect of these three factors entirely determines the time it takes for the animals to successfully solve the lockbox task. This underscores the comprehensiveness of our analytical approach, even though this breakdown into factors isn't the only possible one.

\subsection{Time to Solution ($T$)}

\begin{figure}\center
\includegraphics[width=\linewidth]{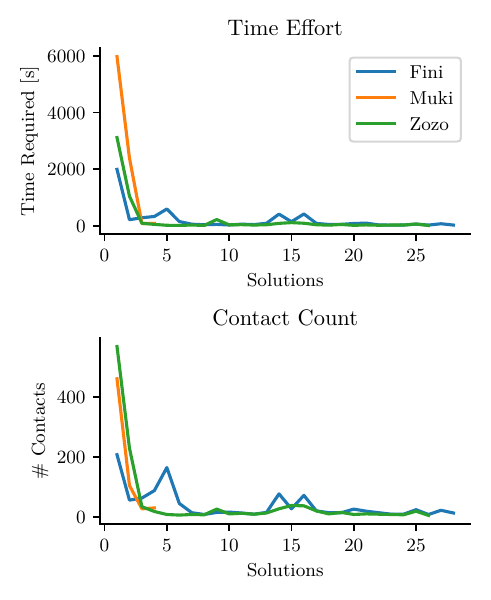}
\caption{Duration in seconds per individual. Plotted over consecutive solutions, each plot shows (a) the time needed, or (b) the amount of contacts needed between consecutive solutions. We can see that in both measures all three birds quickly adapt and require much less effort after their first or second solution.}
\label{fig:opening_effort}
\end{figure}

Our initial step in the analysis involves quantifying task performance, which we measure as "time-to-solution" ($T$) – essentially, the time it takes to successfully complete the lockbox task. In Fig.~\ref{fig:opening_effort} we plot the $T$ values for each bird across consecutive solutions, alongside the number of actions performed with the corresponding time frame. Fini and Zozo managed to solve the lockbox in less than an hour during their first attempts, while Muki took about one and a half hours. However, after the first success the time to solve the problem dropped drastically for, and differed little between, all birds. The plots for number of actions show a similar pattern. This rapid reduction in the effort required to solve the task is a distinctive feature of these learning curves. We attribute this reduction to joint adaptation in engagement, mechanical skill, and strategy. To verify this, we will next examine each of these factors individually. 

%
%

\subsection{Mistargeting of Actions ($M$)}

As a first factor that may influence $T$, we analyze the birds' capacity to choose actions that effectively contribute to solving the lockbox. In principle, birds could utilize a wide range of actions as long as these actions fulfil the necessary steps to open the lockbox (align T-bar with slot in wheel, remove wheel by pulling it over the T-bar, push bar to the side, open door). Creating a precise model of these actions would be challenging, as it would involve distinguishing between pushing, pulling, and other actions on the same object. Thus, we evaluate the quality of action selection by comparing the number of contacts the birds make in functional  versus non-functional locations.

Specifically, at the start of the first session and after each consecutive solution, the ‘correct' behavior is to direct all actions to the wheel until it is removed, since it is the only part of the setup that is susceptible to change and functionally the initial component on the path to solution. This means that one measure of learning could be the proportion of total actions addressed to the wheel in the initial state. In other words, we assess the proportion of instances in which the bird did what an ideal intelligent actor would have done. We define $n_{w|w}$ as the number of actions a bird performed on the wheel (left subscript) in the state when the wheel (right subscript) needs to be removed, and define $n_{a|w}$ as the total number of all actions the bird performed in that state. Similarly we define $n_{b|b}$ as the number of actions on the bar when the bar should be touched, $n_{a|b}$ as the total number of actions in that state, $n_{d|d}$ as the number of actions on the door when the door should be touched, $n_{a|d}$ as the total number of actions in that state. The measure $M = \frac{n_a}{n_c} = \frac{n_{a|w}+n_{a|b}+n_{a|d}}{n_{w|w} +  n_{b|b} + n_{d|d}}$ then captures the degree to which the birds make contact in non-functional locations. We call this measure $M$ Mistargeting-of-Actions. Its reciprocal, $p = 1/M$ would represent the probability to make contact with a functionally relevant part of the lockbox. Regarding allocation, a value of $M=1$ indicate that the bird exactly follows the right allocation strategy, while $M \gg 1$ indicates that the bird rarely makes contact with those parts of the lockbox it should move.

\begin{figure}
\includegraphics[width=\linewidth]{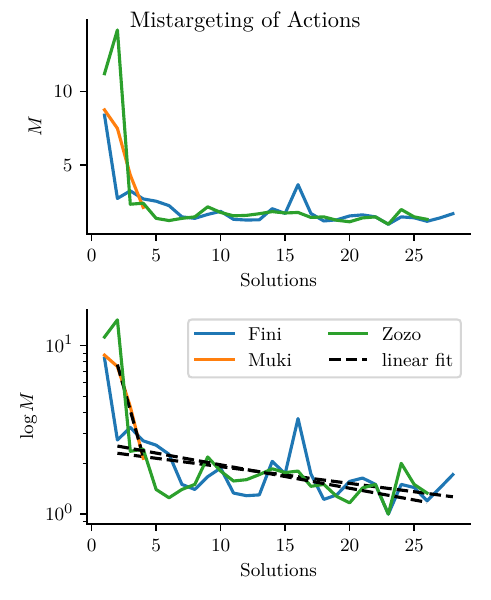}
\caption{Mistargeting of Actions $M$ (left) and conditional probabilities $p(did|should)$ (right); The y-axis values in these plots are the reciprocal of one another. $M$ shows that the birds quickly improve already in their first or second solutions. The second plot, $p(did|should)$ shows that there is an ongoing improvement happening until \~solution 8, for Fini and Zozo.
}
\label{fig:mta}
\end{figure}

Fig.~\ref{fig:mta} shows how $M$ and $log(M)$  evolve as a function of the number of solutions. We also plot $log(M)$ as it helps to highlight adaptation happening during later solutions. $M$ shows that the birds significantly improve in the first one or two solutions. This makes $M$ an important exploratory factor for the birds' improvement in time-to-solution ($T$). The plot of $log(M)$ reveals that there is ongoing adaptation also in later stages of the experiment. For all data \emph{after the first solution}, we fit linear functions to $log(M)$ (the functions are linear in log space). The falling slopes of these linear models indicate that also after the first, initial drop in $M$, there is ongoing adaptation.

\begin{figure*}
\includegraphics[width=\textwidth]{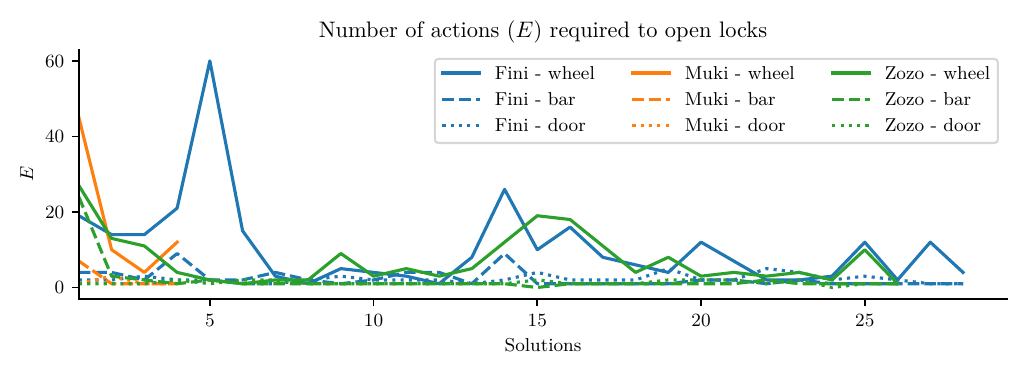}
\caption{Number of correctly targeted actions required to reach consecutive solutions. This plot shows the number of actions required to open a lock, given that lock should be opened at the moment. For all birds, opening the wheel generally requires most actions to be solved. For Muki and Zozo the number of actions required to open the wheel is highest in their first solutions and largely drops from there on. Fini's proficiency at opening the wheel also seems to follow a trend towards less actions needed, however there are two peaks corresponding to the 5th and 14th solutions where she needed to dedicate additional effort.}
\label{fig:skill}
\end{figure*} 

\subsection{Manipulation Effort (E)}

A second factor that might impact time-to-solution is the birds' sensorimotor expertise at removing the wheel.  We quantify this through the amount of correctly addressed actions $n_w$ required to remove the wheel, $n_b$ required to open the bar, and $n_d$ required to open the door, as depicted in Fig.~\ref{fig:skill}. The figure reveals that as the birds progress through the task in subsequent attempts, they require fewer actions directed towards the wheel to complete that stage compared to their earlier sessions. The total number of functionally relevant contacts $n_c$ accumulates as the sum $n_w+n_b+n_d$.

It is worth noting specific observations from individual sessions. The data for Fini, as shown in Fig.~\ref{fig:skill}, differ from those of the other two birds in that this subject shows two sessions ( 5th and 14th) where the number of wheel-directed actions used to remove the wheel was significantly larger than expected for that level of expertise (Video). In these sessions Fini displaced the wheel forward, but without rotating it sufficiently. As a result, the slit was misaligned to the pin, preventing the wheel from being fully removed. We cannot distinguish whether this was simply an ‘error' or an exploratory variation in behavior, but such events are likely part of the learning process: the birds need to fail in order to learn that both forward displacement and rotation are necessary for success.

\subsection{Engagement / Inter-Contact Interval ($\Delta$)}

A third factor that influences time-to-solution is the level of ‘engagement' of animals with the task posed to them. Often, an animal's performance in a task is not solely affected by its competence but also by ‘motivational' factors, somewhat akin to the differentiation between competence and performance in linguistic behavior~\cite{chomsky_aspects_1965}. Occasionally, an individual that has previously demonstrated competence in solving a task may ‘lose interest' and either remain inactive or engage in behavior unrelated to the task at hand. It is likely that such factors evolve alongside the learning process, and neglecting in our analysis could lead to overlooking vital information. In order to avoid such issues we evaluate the animals' engagement with the task.

\begin{figure}\center
\includegraphics[width=\linewidth]{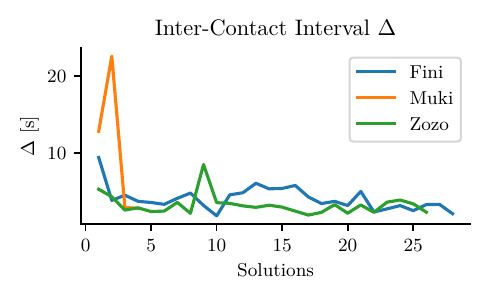}
\caption{Engagement of the birds with the lockbox. Engagement was operationalized as Inter-Contact Interval ($\Delta$), the average time-duration between physical interactions with the lockbox. Lower $\Delta$ corresponds to higher engagement. The birds Fini, Muki, and Zozo start with different levels of engagement and especially Muki and Fini become more engaged over time. Zozo is relatively engaged with the lockbox from the beginning, but also becomes more engaged.}
\label{fig:engagement}
\end{figure}

To assess the birds' engagement with the puzzle, we measure the average time-duration between timestamps where the birds initiate contact with the lockbox. We refer to this measure as Inter-Contact Interval and denote it $\Delta$, where $\Delta=T/n_a$. Lower $\Delta$ values correspond to a lower time to solution, all other factors being equal. In Fig.~\ref{fig:engagement} we plot engagement as a function of the number of solutions. Relative to their first solution, all 3 birds show an increase in engagement (signified by a decrease in $\Delta$) in subsequent solutions, except for one outlier in Zozo's ninth solution. Table~\ref{table:engagement} shows data for the first and last solution of each bird tabularly. It displays $\Delta$ alongside its reciprocal \emph{Rate of work}, as well as the number of contacts used to compute these metrics.

\begin{table}
\small\sf\centering

%

\caption{Increased engagement measured as decrease in \emph{Inter-contact interval $\Delta$}}
\label{table:engagement}

\begin{tabular}{ |c|c|c|c|c| } 
 \hline
  &  Solution & Duration[s] & \# Contacts & $\Delta$\\ \hline
 \multirow{2}{2em}{Fini} & First & 1993.94 & 208 & 9.62\\ \cline{2-5}
  & Last & 27.92 & 13 & 2.10\\ 
  \hline
   \multirow{2}{2em}{Muki} & First & 5978.41 & 462 & 14.29\\ \cline{2-5}
  & Last & 88.554 & 30 &  2.95\\ 
  \hline
   \multirow{2}{2em}{Zozo} & First & 3117.764 & 570 & 5.46\\ \cline{2-5}
  & Last & 11.72 & 5 & 2.34\\ 
 \hline
\end{tabular}

\end{table}

\subsection{Joint Analysis of Factors M, E, $\Delta$}

In order to understand how the animals in our study learn to solve the lockbox, we find it valuable to jointly examine three aspects: strategy, sensorimotor skill, and engagement. This decomposition $T=M * E * \Delta$ is a straightforward mathematical identity, as becomes obvious in Eq.~\ref{eq:decomposition}

\begin{equation}\label{eq:decomposition}
T = n_a * \frac{T}{n_a} = \frac{n_a}{n_c} * n_c * \frac{T}{n_a} = M * E * \Delta
\end{equation}

However, beyond this straightforward mathematical reformulation, there are compelling arguments for adopting this decomposition. Intelligent behavior arises from a combination of heterogeneous processes that interact and rely on each other. The three factors we propose as an explanation for task-performance may serve as examples of such heterogeneous, interdependent processes. An agent cabable of solving the lockbox requires a minimum level of competence in action selection strategy and mechanical skill, in addition to being engaged with the puzzle. The multiplicative relation that results in $T$ is also logically consistent with the insight that the bird can divide $T$ by half either by learning to remove the wheel with half as many actions (adaptation in $E$), by mistargeting half as many actions (adaptation in $M$), or by working on the puzzle twice as fast (adaptation in $\delta$). These are extreme examples, combinations thereof are also possible.

\begin{figure*}\center
\includegraphics[width=\linewidth]{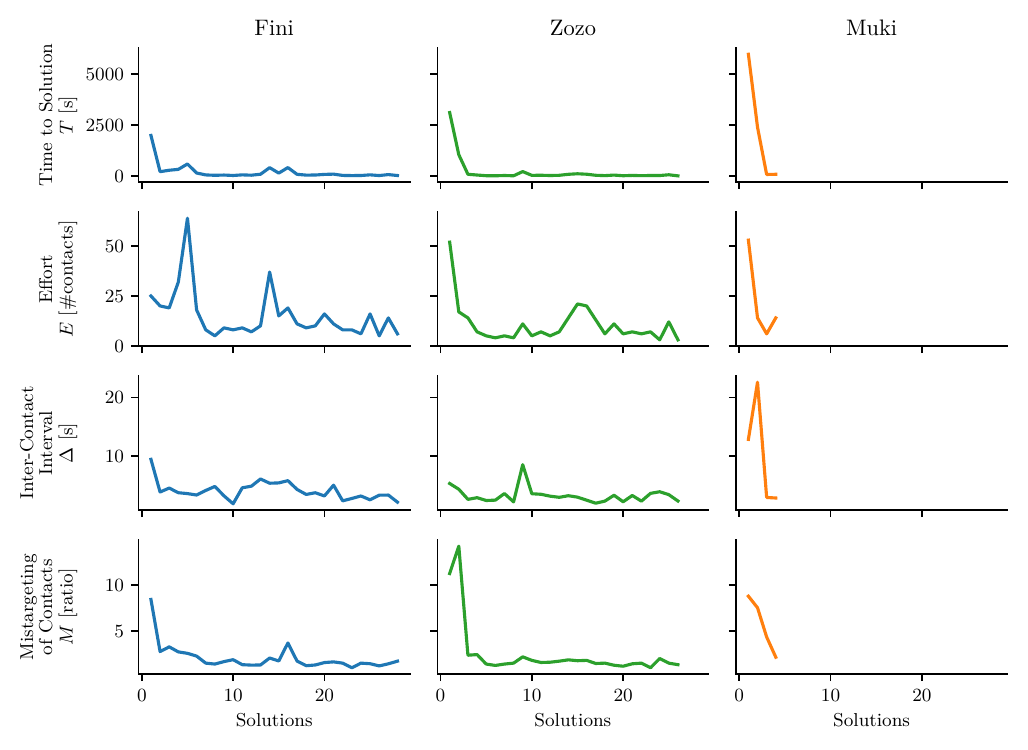}
\caption{Task performance and three different factors that may contribute to it, plotted individually per bird. For all three birds, the task-performance improves substantially after their first one or two solutions. But different combinations of factors may explain that improvement.}
\label{fig:summary}
\end{figure*}

For each bird, these three factors and $T$ evolve differently over time. This is evident in Fig.~\ref{fig:summary}, which provides a visual summary of these measurements. In all three birds, $T$ decreases to low values after a few solutions, with the most significant drop in $T$ occuring between the first and second solution. The data underscores that none of the three candidate factors can singly explain adaptation for all three birds. Fini demonstrates the most noteworthy change in $\Delta$ and $M$, while Zozo shows most change in $E$ and $M$, while Muki shows relevant decreases in $E$, $\Delta$, and $M$.  This supports our view that an analysis of adaptation in challenging tasks should strive to encompass multiple of the involved adaptive processes. It would not be possible to reconstruct $T$ solely from just one of the analyzed factors.

\section{Constraining the Space of Mechanistic Models}\label{sec:constraints}

Goffin's cockatoos are complex animals that exhibit intricate behavior. This is especially the case in environments that offer many affordances, as is the case for the lockbox. Given this complexity, it is not promissing to try to derive a single, highly detailed mechanistic explanation from limited data. However, a more attainable goal is to instead extract constraints on possible mechanisms. Our proposed strategy is to build up an increasing set of such constraints across multiple studies. Importantly, these constraints should not be purely descriptive, they need to be formulated in the space of mechanistic models.

In the following, we will relate the descriptive results from the previous section to properties of mechanisms that could account for them. We will explore how the observations we've made can render certain mechanistic explanations more plausible than others.

\subsection{Constraint A – Adaptation in Multiple Interacting Factors}

Our model, $T=E*M*\Delta$, represents a decomposition of the birds' task performance for each individual solution. The data illustrates that not only does $T$ decrease for each bird from the first to the last solution, but each of the three factors also individually decreases (non-monotonously) for each bird. 

If we aim to utilize an algorithmic model to elucidate the evolution of the birds' time to solution, such a model should provide an explanation for the adaptation observed in each of those factors. Additionally, a robust model should also propose a hypothesis of how these factors interact. While each factor contributes linearly to $T$ for a single task solution, it is likely that these factors interact and evolve non-linearly over the course of the experiment. This does not necessarily mean that a mechanistic model must implement that factorization, but an analytical treatment of that mechanistic model must yield similar results as the one we have observed in the birds.

\subsection{Constraint B – Slow and Fast Adaptation}

Our observations reveal that the birds exhibited both rapid and gradual adaptation in various aspects of their behavior. Specifically, the performance in the task, measured by time-to-solution $T$, showed a significant and swift improvement in the initial few solutions for all birds.  This steep decline in $T$ can be explained by steep decline in several of our proposed factors. But in addition to these steep changes in several measures, we can also observe slow, long-term adaptation in the birds' capability to target functional parts of the lockbox (decreasing $M$). 

An algorithmic model should allow for fast adaptation in each of the factors we proposed, even based on just a single successful trial. However, an effective model should also enable the agent's behavior to steadily adapt over the course of many sessions. The fast adaptation based on only a few trials indicates that established reinforcement learning models may not be a good explanation on their own, as these typically require many trials to significantly improve an agent's performance.

\subsection{Constraint C – Regress to Old Solution Behavior and Non-monotony}

Two out of the three birds (Muki and Zozo) display a temporary decline in wheel-directed actions after their initial success. This represents the most notable example of a non-monotonic adaptive pattern in our data. Additionally we observe that other variables also do not follow a strictly monotonous increase or decrease.

An algorithmic model should be able to replicate such re-emergence of behaviors that were temporarily suppressed. Such re-emergence could potentially be explained by a reinforcement learning agent's erroneous credit assignment. As the birds indeed had to touch the door and bar prior to receiving the cashew reward, a discounted reinforcement scheme could be the driving mechanism behind this re-emergence. Interestingly, this constraint appears to be in conflict with Constraint B. Resolving this apparent conflict will likely be insightful.

\subsection{Constraint D – Old Strategy on New Lockbox}

At the outset of the experiment, the birds primarily acted on the new lockbox (with wheel) as if it was the old lockbox from the habituation training (without the wheel). Only when this behavior did not lead to movement or success, they started to exhibit more varied behavior, including exploration of the wheel. This initial tendency of the birds to apply the strategy they learned during the habituation phase suggests that they may not perceive a significant alteration in the lockbox sequence or kinematics.

An algorithmic model should be able to replicate this behavior. This implies that models involving a high degree of physical or kinematic reasoning, which might lead an agent to immediately recognize that the wheel obstructs the motion of the bar, are less likely to be applicable. It also indicates that mechanisms with sensitive novelty detection on the appearance of the lockbox are improbable explanations.

\subsection{Constraint E –  Inter-Individual Differences}

The individuals in our study adapt in different ways to solve the task. For instance, the steep decrease in time-to-solution for Muki can be primarily attributed to an improvement in mechanical skill, whereas for Fini, the most suitable explanation for this decrease involves a combination of changes in targeting of actions and engagement. 
An algorithmic model must either incorporate enough randomness in its adaptation process, or in its initialisation to allow for such distinctive adaptive trajectories. 

\subsection{Using Constraints}

In order to identify the mechanisms underlying behavior, we need to infer those mechanisms from data. However, it is unlikely that a single experiment, or even a limited number of experiments, can provide sufficient information to infer the complex, interacting processes giving rise to situated manipulation behavior. As there is not enough data available for such inference, currently and in the foreseeable future, we will have to establish incremental methods to make coarse inference about mechanisms. By outlining the above constraints, our aim is to start such a coarse inference process, directed towards iteratively identifying the mechanisms underlying Goffin's cockatoos mechanical problem solving.

Such an approach requires us to link high-level, abstract descriptions of biological behavior with high-level abstract properties of artificial models. Over many years, behavioral biology has extracted many such properties to describe biological behavior. However, although many algorithmic models of behavior have been developed in robotics and related disciplines, these models usually do not receive a thorough analytical treatment, analogous to behavioral research in biology. To foster the described constraint-based approach, it would be required to largely apply biological research methods to artificial systems, like robots. Similar proposals have previously been made in~\cite{lazebnik_can_2002} and~\cite{jonas_could_2017}, however not with the intent to promote a constraint-based approach to identifying proximate mechanisms of behavior.

\section{Conclusion}

To understand acquisition of problem solving skills, we need to test animals in challenging tasks, such as the lockbox. Here, adaptation and solution behavior recruit many heterogeneous, interacting processes in the animal. This makes behavior generation complex in the sense that we need to jointly analyze these processes to understand behavior and its adaptation. Thus, in this study we have  jointly analyzed different factors that change as Goffin's cockatoos learn to solve lockboxes. 

We analyzed the animals' engagement, mechanical skill and targeting of contact actions, and were able to show that each of these factors impacts time-to-solution in our experiments. Each factor contributed differently between animals and also for different solutions of the same animal. This highlights and confirms that a joint analysis of the involved factors is crucial, if we aim to comprehensively explain behavioral adaptation in such challenging tasks.

The intricate nature of the involved processes and their complex interaction makes it unlikely to formulate a plausible explanatory model from limited data in a single study. Therefore, we advocate for a constraint-based approach, where we relate abstract properties of behavioral data to properties of algorithmic mechanisms. This constitutes an inference problem that is attainable with less data, as it aims to infer less detailed properties of the target behavior.

The constraints we have identified in this study are not as detailed as a full algorithmic model would be. However, they represent hypotheses related to proximate mechanisms. Therefore, they are a starting point to bridge between the available data and explanatory mechanical models. As this study is explorative, these constraints should be seen as initial hypotheses that should be tested in future experiments.

To understand adaptive behavior in challenging scenarios, such as the lockbox, we need to develop suitable research methodology. The constraint-based approach is an initial step to develop such methods, with the goal to iteratively build mechanistic models for complex behavior. We hope that other researchers find such an iterative approach to modeling intriguing and that we can spark a further development in research methods to build mechanical models of animal behavior.

\section{Acknowledgements}
We are very thankful to Alex Kacelnik for his fruitful feedback and intellectual contributions. Alex was a driving force behind this paper, instrumental at all stages of paper writing and it would not have been possible to write this paper without him. We are further thankful to Huu Duc Nguyen for his feedback on the paper, as well as Lukas Schattenhofer, Max Winkelmann and Rory Michele assisting in data analysis.

\bibliography{references}

\end{document}